\documentclass[12pt,fleqn]{article} 
 
\begin{document} 
\begin{titlepage}
\vspace*{-1.2in}
\centerline{\hfill                 IASSNS-HEP-99/109}\bigskip

\begin{center}

{\large \bf Spacetime metric from linear electrodynamics II} 
\\
\bigskip
\bigskip
\bigskip
\centerline{Friedrich W.\ Hehl$^{\dagger,{\$},}$\footnote{email: 
{\tt hehl@ias.edu, hehl@thp.uni-koeln.de}},
Yuri N.\ Obukhov$^{\ast,}$\footnote{email: {\tt yo@thp.uni-koeln.de}},
and Guillermo F.\ Rubilar$^{{\$},}$\footnote{email: {\tt
gr@thp.uni-koeln.de}}}
\bigskip
\bigskip

\centerline{$^\dagger$School of Natural Sciences}
\centerline{Institute for Advanced Study}
\centerline{Princeton, NJ 08540, USA}
\bigskip\bigskip

\centerline{$^\ast$Department of Theoretical Physics}
\centerline{Moscow State University}
\centerline{117234 Moscow, Russia}
\bigskip\bigskip

\centerline{$^{\$}$Institute for Theoretical Physics}
\centerline{University of Cologne}
\centerline{50923 K\"oln, Germany}
\bigskip\bigskip


{\bf Abstract}

\end{center}
\begin{quotation}
Following Kottler, \'E.Cartan, and van Dantzig, we formulate the
Maxwell equations in a metric independent form in terms of the field
strength $F=(E,B)$ and the excitation $H=({\cal D}, {\cal H})$. We
assume a linear constitutive law between $H$ and $F$. First we split
off a pseudo-scalar (axion) field from the constitutive tensor; its
remaining 20 components can be used to define a duality operator $^\#$
for 2-forms. If we enforce the constraint $^{\#\#}=-1$, then we can
derive of that the conformally invariant part of the {\em metric} of
spacetime. {\em file weimar16a.tex, 1999-11-24}
\end{quotation}
\vfill
\end{titlepage}

\pagebreak

\section{Axiomatics of metric-free electrodynamics} 

We assume that spacetime is described by a smooth 4-dimensional
manifold $X$ and that it is possible to foliate $X$ into 3-dimensional
submanifolds which can be numbered by a monotonously increasing
parameter $\sigma$. This could be called Axiom 0.
 
Our formulation of classical electrodynamics \cite{PLH,OH,HOR} is
based on four axioms, three of which being of general validity and
independent of the metric and/or affine structures of spacetime. Only
in the context of the fourth axiom, the constitutive relation, the
metric comes into play.

\subsection{Axiom 1: Electric charge conservation} 
 
We assume the existence of a conserved electric current described by
means of an odd 3-form $J$ on $X$; for the exterior calculus involved,
see \cite{Ted}. Conservation of electric charge is a firmly
established fact which basically can be verified by {\em counting}
charged elementary particles inside a closed region.  Mathematically,
$J$ is conserved when
\begin{equation} \label{axiom1}  
 \oint\limits_{\Omega_3}J=0\,,\qquad \partial \Omega_3=0\, , 
\end{equation}   
where $\Omega_3$ is an arbitrary closed 3-dimensional submanifold of
the 4-manifold $X$. By de Rham's theorem, the current $J$ is not only
closed $dJ=0$, but also exact, see \cite{Toupin,Post95}. Thus the
inhomogeneous Maxwell equation is a consequence of (\ref{axiom1}),
\begin{equation}\label{ime} J=dH\, ,\end{equation}  
with $H$ as the odd electromagnetic {\em excitation} 2-form.  Note
that $H$ has an independent {\em operational} interpretation, see
\cite{HOR}.

\subsection{Axiom 2: Existence of the Lorentz force density} 
 
We introduce a field of frames $e_\alpha$ as reference system in $X$;
by Greek letters $\alpha,\beta,\dots=0,1,2,3$, we denote anholonomic
or frame indices.  The odd current 3-form $J$, together with the force
density $f_\alpha$ (odd covector-valued 4-form), the notion of which
is assumed to be known from mechanics, allows us to formulate the
Lorentz force density as
\begin{equation} \label{axiom2} 
  f_\alpha= (e_\alpha\rfloor F) \wedge J\,. 
\end{equation}   
Thereby the electromagnetic field strength $F$ is defined as an even
2-form.

\subsection{Axiom 3: Magnetic flux conservation} 
  
It is possible to count single quantized magnetic flux lines inside
superconductors of type II. This suggests to take the conservation of
magnetic flux as axiom 3,
\begin{equation} 
  \oint\limits_{\Omega_2}F=0\,,\quad \partial \Omega_2=0\,,
\end{equation} 
for an arbitrary closed submanifold $\Omega_2$. As a consequence, we
find the homogeneous Maxwell equation
\begin{equation}\label{hme} dF=0\,,\end{equation}  
and the exactness of $F$, i.e., $F=dA$.

The Maxwell equations (\ref{ime}) and (\ref{hme}) are automatically
diffeomorphism invariant, are independent of metric and connection
(like the exterior {\em and} interior products), and are valid in this
form in special and general relativity in arbitrary frames and in the
post-Riemannian spacetimes of gauge theories of gravity, see
\cite{PLH}.

\subsection{Axiom 4: Constitutive law}  
 
To complete the formulation, a relation between $H$ and $F$ (and
possibly $J$) is required, namely the constitutive law.  We confine
ourselves to a {\em linear} constitutive law (`linear
electrodynamics'). In terms of the components of $H$ and $F$ in an
arbitrary coordinate system $x^i$, it reads ($i,j\dots=0,1,2,3$),
\begin{equation}\label{cl} 
H_{ij}=\frac{1}{4}\,\epsilon_{ijkl}\ \chi^{klmn}\ F_{mn}, 
\qquad{\rm with}\qquad {\chi}^{ijkl}=-{\chi}^{jikl}=-{\chi}^{ijlk}\,,
\end{equation} 
where $\epsilon_{ijkl}$ is the Levi-Civita symbol and $\chi^{ijkl}$ an
even tensor density of weight $+1$. Taking into account that the free
field Lagrangian is of the form $L\sim H\wedge F$, we find
${\chi}^{ijkl}={\chi}^{klij}$, leaving only 21 independent functions
for ${\chi}^{ijkl}$. 

Moreover, one can decompose $\chi^{ijkl}$ according to
\begin{equation}\label{decomp}  
 {\chi}^{ijkl}=f(x)\,\stackrel{\rm o}{\chi}{}^{ijkl} +\alpha(x)\,
\epsilon^{ijkl} \,,\qquad{\rm with}\qquad \stackrel{\rm
o}{\chi}{}^{[ijkl]}\equiv 0\, .
\end{equation} 
Here $f(x)$ is a dimensionfull scalar function such that
$\stackrel{\rm o}\chi{}^{ijkl}$ is dimensionless. The pseudo-scalar
constitutive function $\alpha(x)$ can be identified as an Abelian
axion field. It was first discussed by Ni \cite{Ni73,Ni77,Ni99}. Its
experimental bounds have been discussed by Carroll et
al.\cite{Carroll}. Usually, however, an hypothetical axion is
considered in the context of {\em non}-Abelian gauge theories, see
\cite{Weinberg,Wilczek1,Moody}.  Note that ${\stackrel{\rm o}
{\chi}}{}^{ijkl}$ has analogous algebraic symmetries and the same
number of 20 independent components as a Riemannian curvature tensor:
\begin{equation}
\stackrel{\rm o}{\chi}{}^{ijkl}= -\stackrel{\rm
o}{\chi}{}^{jikl}=-\stackrel{\rm o}{\chi}{}^{ijlk}= \stackrel{\rm
o}{\chi}{}^{klij}\,,\qquad \stackrel{\rm o}{\chi}{}^{[ijkl]}=0\,.
\end{equation}

\section{Duality operator ${}^\#$ and its closure} 
  
By means of axiom 4, a new duality operator can be defined acting on
2-forms on $X$. In components, an arbitrary 2-form
$\Theta=\frac{1}{2}\Theta_{ij}\, dx^i\wedge dx^j$ is mapped into the
2-form ${}^\#\Theta$ by
\begin{equation}\label{dual} 
 {}^\#\Theta_{ij}:=\frac{1}{4}\epsilon_{ijkl}\,  
 {\stackrel{\rm o}{\chi}}{}^{klmn}\, \Theta_{mn}\, . 
\end{equation} 
No metric is involved in this process. Now the linear law
(\ref{cl}) can be written as
\begin{equation} 
H=\left(f\,{}^\# +\alpha \right)F. 
\end{equation} 
 
We postulate that the duality operator, applied twice, should, up to a
sign, lead back to the identity. By this closure relation or the
`electric and magnetic reciprocity' \cite{Toupin}, we can
additionally constrain the 20 independent components of $\stackrel{\rm
o}{\chi}$ without using a metric. This appears to be a sufficient
condition for the {\em non}existence of {\em birefringence} in vacuum,
see \cite{Ni73,Ni77,Claus,Ni99,HLaem,Laemm,Fukui}. Therefore, we
impose
\begin{equation}\label{cr} 
 {}^{\#\#}=-1 \, .  
\end{equation}  
The minus sign yields Minkowskian signature\footnote{A different
duality operator could be defined by ${}^{\widehat\#}\Theta_{ij}=
\frac{f}{4}\epsilon_{ijkl}\,{\stackrel{\rm o}{\chi}}{}^{klmn}\,
\Theta_{mn}$. It would satisfy ${}^{\widehat{\#}\widehat{\#}}=-f^2$.},
whereas the condition $^{\#\#}= +1$ would lead to Euclidean or to the
mixed signature $(+,+,-,-)$.
 
Seemingly Toupin \cite{Toupin} and Sch\"onberg \cite{Schoenberg} were
the first to deduce the conformally invariant part of a metric from
relations like (\ref{dual}) and (\ref{cr}). This was rediscovered by
Jadczyk \cite{Jadczyk}, whereas Wang \cite{Wang} gave a revised
presentation of Toupin's results. A forerunner was Peres \cite{Peres},
see in this context also the more recent papers by Piron and Moore
\cite{Piron75,Piron95}.  It was recognized by Brans
\cite{Brans71,Brans74} that, within general relativity, it is possible
to define a duality operator in much the same way as presented above,
see (\ref{dual}) and (\ref{cr}), and that from this duality operator
the metric can be recovered. Such structures were subsequently
discussed by numerous people, by Capovilla et al.\ \cite{Capo89}, 't
Hooft \cite{tHooft91}, Harnett \cite{Harnett1}, and Obukhov \&
Tertychniy \cite{Tert}, amongst others, see also the references given
there.

It is convenient to adopt a more compact {\it bivector} notation by
defining the indices $I,J,\dots = (01, 02, 03, 23, 31, 12)$. Then
${\stackrel{\rm o}{\chi}}{}^{ijkl}$ becomes the symmetric $6\times 6$
matrix $\stackrel{\rm o}{\chi}{}^{IK}$ and (\ref{cr}) reads
\begin{equation}\label{cr2} 
 \stackrel{\rm o}{\chi}{}^{IJ}\,\epsilon_{JK}\stackrel{\rm
o}{\chi}{}^{KL}\, \epsilon_{LM}=-\delta^I_M \,,\qquad{\rm with}\qquad
\epsilon_{IJ} \stackrel{\rm o}{\chi}{}^{IJ}\equiv 0\, .
\end{equation} 
In terms of $3\times 3$-matrices
\begin{equation}  
 \stackrel{\rm o}{\chi}{}^{IJ}=\stackrel{\rm o}{\chi}{}^{JI}=  
 \left(\begin{array}{cr}A& C \\C^{{\rm T}} &B 
 \end{array}\right),\quad \epsilon^{IJ} =\epsilon^{JI}=  
 \left(\begin{array}{cr}0& {\mathbf 1} \\{\mathbf 1} 
&0\end{array}\right)\, , 
\end{equation}  
where $A=A^{\rm T}\,,B=B^{\rm T}$, and the superscript ${}^{\rm T}$
denotes transposition. The general non-trivial solution of the closure
relation (\ref{cr2}) is
\begin{equation}\label{??} 
 \stackrel{\rm o}{\chi}{}^{IJ}=\left(\begin{array}{cc}pB^{-1} + qN& 
B^{-1}K \\  
 -KB^{-1} &B \end{array}\right)\, . 
\end{equation}  
Here $B$ is a nondegenerate arbitrary {\em symmetric} $3\times 3$
matrix (6 independent components $B_{ab}$), $K$ an arbitrary {\em
antisymmetric} matrix (3 independent components $K_{ab}=:
\epsilon_{abc}\, k^c$), $N$ the symmetric matrix with components
$N^{ab}:=k^ak^b$, and $q:= - 1/{\det B}$, $p:=[{\rm tr}(NB)/\det B] -
1$. Thus, Eq.(\ref{??}) subsumes 9 independent components.

\section{Selfduality and a triplet of 2-forms}

The duality operator ${}^\#$ induces a decomposition of the
6-dimensional space of 2-forms into two 3-dimensional invariant
subspaces corresponding to the eigenvalues $\pm i$. The 2-form basis
$dx^i\wedge dx^j$ or $\Theta^I$ decomposes into two 3-dimensional
column vectors
\begin{equation} 
 \Theta^I = \left(\begin{array}{c} \beta^a \\ \gamma_b 
 \end{array}\right)\,,\quad a,b, \dots = 1,2,3. 
\end{equation} 
The self-dual basis ${\stackrel {({\rm s})} \Theta}{}^I =
\frac{1}{2}(\Theta^I - i\,{}^\#\Theta^I)$ decomposes similarly into
\begin{equation} 
{\stackrel {({\rm s})} \Theta}{}^I = \left(\begin{array}{c}  
{\stackrel {({\rm s})} \beta}{}^a \\  
{\stackrel {({\rm s})} \gamma}_b \end{array}\right)\,. 
\end{equation}    
One of the 3-dimensional invariant subspaces can be spanned by, say,
${\stackrel {({\rm s})} \gamma}$. Then ${\stackrel {({\rm s})} \beta}$
can be expressed in terms of ${\stackrel {({\rm s})} \gamma}$
according to ${\stackrel{({\rm s})}\beta}=(i + B^{-1}K)B^{-1}
{\stackrel{({\rm s})}\gamma}$.  Therefore ${\stackrel {({\rm s})}
\gamma}$ or,
equivalently, the {\em triplet of 2-forms}
\begin{equation}  
 S^{(a)}:= -(B^{-1})^{ab}\,\,{\stackrel{({\rm s})}{ \gamma}}{}_b\,, 
\end{equation}  
subsume the properties of this invariant subspace. Each of the 2-forms
carry 3 independent components, i.e., they add up to 9 components.
 
The information of the constitutive matrix $\stackrel{\rm o}
\chi{}^{IJ}$ is now encoded into the triplet of 2-forms $S^{(a)}$.
The latter satisfy the completeness relation
\begin{equation}\label{cc} 
 S^{(a)}\wedge S^{(b)} = {\frac 1 
3}\,(B^{-1})^{ab}\,(B)_{cd}\,S^{(c)}\wedge S^{(d)} \, . 
\end{equation}

\section{Extracting the metric} 
 
Within the context of $SU(2)$ Yang-Mills theory, Urbantke
\cite{Urban1,Urban2} was able to derive a 4-dimensional metric
$g_{ij}$ from a triplet of 2-forms satisfying a completeness condition
of the type (\ref{cc}).  This procedure also applies in our
$U(1)$-case. Explicitly, the Urbantke formulas read
\begin{eqnarray}
  \sqrt{{\det}\,g}\;g_{ij} & =& -\,{\frac 2 3}\,\sqrt{\det 
    B}\,\epsilon_{abc}\, \epsilon^{klmn}\,S^{(a)}_{ik}S^{(b)}_{lm} 
  S^{(c)}_{nj}\,, \label{uf1}\\ \sqrt{{\det}\,g} & =& -\,{\frac 1 
    6}\,\epsilon^{klmn}\,B_{cd}\, 
  S^{(c)}_{kl}S^{(d)}_{mn}\, \label{uf2}.
\end{eqnarray} 
The $S^{(a)}_{ij}$ are the components of the 2-form triplet $S^{(a)} =
S^{(a)}_{ij} dx^i\wedge dx^i/2$.  If we substitute the $S^{(a)}$ into
(\ref{uf1}) and (\ref{uf2}), we can display the metric explicitly in
terms of the constitutive coefficients:
\begin{equation}\label{metric} 
g_{ij} = {\frac 1 {\sqrt{\det B}}}\left(\begin{array}{c|c} \det B & 
-\,k_a \\ \hline -\,k_b & -\,B_{ab} + (\det B)^{-1}\,k_a\,k_b 
\end{array}\right) \, .
\end{equation}  
Here $k_a:=B_{ab}k^b = B_{ab}\,\epsilon^{bcd}K_{cd}/2$. One can
verify that the metric in (\ref{metric}) has Minkowskian signature.
Since the triplet $S^{(a)}$ is defined up to an arbitrary scalar
factor, we obtain a {\em conformal} class of metrics.

\section{Properties of the metric} 
 
\subsection{Hodge duality operator} 
 
The inverse of (\ref{metric}) is given by 
\begin{equation}\label{metricinv} 
 g^{ij}= \frac{1}{\sqrt{\det B}}\left(\begin{array}{c|c}  
  1- (\det B)^{-1} k_c k^c & -k^b\\ \hline  
  -k^a & -(\det B)(B^{-1})^{ab}\end{array} \right)\, . 
\end{equation} 
With the help of (\ref{metric}) and (\ref{metricinv}), we can define
the Hodge duality operator ${}^\star$ attached to this metric. In
terms of the components of the 2-form $F$, we have
\begin{equation} 
^\star F_{ij}:=\frac{\sqrt{-g}}{2}\,\epsilon_{ijkl}\,
g^{km}g^{ln}F_{mn}.
\end{equation} 
This equation can be rewritten, in analogy to (\ref{dual}), by
defining the constitutive tensor $\stackrel{\rm g}\chi{}^{ijkl}$:
\begin{equation}  
 ^\star F_{ij}=\frac{1}{4}\epsilon_{ijkl}\, \stackrel{\rm
 g}\chi{}^{klmn} F_{mn}\, , \quad{\rm with}\quad \stackrel{\rm
 g}\chi{}^{ijkl}:=\sqrt{-g}\left( g^{ik}g^{jl}-g^{jk}g^{il}\right)\, .
\end{equation} 
Note that $\stackrel{\rm g}\chi{}^{ijkl}$ is invariant under conformal
transformations $g_{ij}\rightarrow e^{\lambda(x)}g_{ij}\,$; this takes
care that only 9 of the possible 10 components of the metric can ever
enter $\stackrel{\rm g}\chi{}^{ijkl}$.

In order to compare $\stackrel{\rm g}\chi{}^{ijkl}$ with the
 constitutive matrix $\chi^{IJ}$, we put it in the form
\begin{equation}  
 \stackrel{\rm g}\chi{}^{IJ}= \left( \begin{array}{c|c} \stackrel{\rm
 g}A & \stackrel{\rm g}C \\ \hline \stackrel{\rm g}C{}^{\rm T} &
 \stackrel{\rm g}B\end{array} \right).
\end{equation} 
Then straightforward calculations yield
\begin{eqnarray} 
 \stackrel{\rm g}A{}^{ab}&=&g^{00}g^{ab}-g^{0a}g^{0b}
              =p\,(B^{-1})^{ab}+q\, N^{ab}=A^{ab}\, , \\ \stackrel{\rm
              g}B_{ab}&=&\frac{1}{4}\left(g^{ce}g^{df}-g^{de}g^{ef}\right)
              \epsilon_{acd}\, \epsilon_{efb}=B_{ab}\, , \\
              \stackrel{\rm g}C{}^{a}{}_b&=&\frac{1}{2}\left(
              g^{0c}g^{ad}-g^{ac}g^{0d} \right) \epsilon_{bcd}=
              (B^{-1})^{ad} K_{db}=C^{a}{}_b.
\end{eqnarray} 
Thus, $\stackrel{\rm g}\chi{}^{IJ}=\;\stackrel{\rm o}\chi{}^{IJ}$,
i.e., {the metric extracted allows us to write the original duality
operator $^\#$ as Hodge duality operator}, $^\#={}^\star$.  Therefore,
the original constitutive tensor (\ref{decomp}) can then be written as
\begin{equation} 
 {\chi}^{ijkl}=f(x)\,\sqrt{-g}\left(g^{ik}g^{jl}
 - g^{jk}g^{il}\right) + \alpha(x)\,\epsilon^{ijkl}\,.\label{decomp1} 
\end{equation} 
This representation naturally suggests an interpretation of $f(x)$ as
a scalar {\em dilaton} type field.\footnote{In the low-energy string
models this factor is usually written as $f(x) = e^{-b\phi(x)}$ with a
constant $b$ and the dilation field $\phi(x)$.}

\subsection{Isotropy} 
 
Given a metric, we can define the notion of local isotropy. Let
$T^{i_1\dots i_p}$ be the contravariant coordinate components of a
tensor field and $T^{\alpha_1\dots\alpha_p} :=e_{i_1}{}^{\alpha_1}
\cdots e_{i_p}{}^{\alpha_p}\, T^{i_1\dots i_p}$ its frame
components with respect to an orthonormal frame $e_\alpha =
e^i{}_\alpha\,\partial_i$.  A tensor is said to be locally isotropic
at a given point, if its frame components are invariant under a
Lorentz rotation of the orthonormal frame. Similar considerations
extend to tensor densities.

There are only two geometrical objects which are numerically invariant
under (local) Lorentz transformations: the Minkowski metric
$o_{\alpha\beta}={\rm diag}(+1,-1,-1,-1)$ and the Levi-Civita tensor
density $\epsilon_{\alpha\beta\gamma\delta}$. Thus
\begin{equation}\label{iso} 
{\cal T}^{ijkl}=\phi(x)\, \sqrt{-g}\left( g^{ik}g^{jl} - g^{jk} g^{il}
\right) + \varphi(x)\, \epsilon^{ijkl}
\end{equation}  
is the most general locally isotropic contravariant fourth rank tensor
density of weight $+1$ with the symmetries ${\cal T}^{ijkl}=-{\cal
T}^{jikl} =-{\cal T}^{ijlk}={\cal T}^{klij}$. Here $\phi$ and
$\varphi$ are scalar and pseudo-scalar fields, respectively.

Accordingly, in view of (\ref{decomp1}), we have proved that the
constitutive tensor (\ref{decomp}) with the closure property
(\ref{cr}) is {\em locally isotropic with respect to the metric}
(\ref{metric}), see also \cite{Ni77} and \cite{SU}.

\section{Outlook}  

Developing the ideas of Kottler-Cartan-van Dantzig and following our
previous work \cite{OH}, we demonstrated that the general structure of
classical electrodynamics is fundamentally independent of metric and
connection.

The pseudo-Riemannian metric arises naturally in the context of {\em
linear} electrodynamics from the duality operator $^\#$ constructed in
terms of the constitutive coefficients and from its closure relation
${}^{\#\#}=-1$.  At the same time, two {\em pre}\hskip0.2mm metric
objects emerge from the constitutive law: the pseudoscalar {\em axion}
field $\alpha(x)$ and the scalar {\em dilaton} type field $f(x)$. They
respect all the axioms of electrodynamics (charge conservation, flux
conservation, etc.).

\vspace*{0.25cm} \baselineskip=10pt{\small \noindent The authors are
indebted to Gernot Neugebauer for the invitation to give a talk at the
Journe\'es Relativistes 99 and to Claus L\"ammerzahl, David Moore,
Wei-Tou Ni, and Jan Post for helpful remarks. FWH is grateful to the
IAS for hospitality and to the VW-Foundation, Hannover for
support. GFR would like to thank the German Academic Exchange Service
(DAAD) for a graduate fellowship (Kennziffer A/98/00829). }


\begin{thebibliography}{99} 
 
\bibitem{Brans71} C.H. Brans, {\em Complex 2-forms representation of
the Einstein equations: The Petrov Type III solutions},
{\sl J. Math. Phys.} {\bf 12} (1971) 1616-1619.
 
\bibitem{Brans74} C.H. Brans, {\em Complex structures and
representations of the Einstein equations}, {\sl J. Math. Phys.} {\bf
15} (1974) 1559-1566.
 
\bibitem{Capo89} R. Capovilla, T. Jacobson, and J. Dell, {\em General
relativity without the metric}, {\sl Phys. Rev. Lett.} {\bf 63} (1989)
2325-2328.

\bibitem{Carroll} S.M. Carroll, G.B. Field, and R. Jackiw, {\em Limits
on a Lorentz- and parity-violating modification of electrodynamics},
{\sl Phys.  Rev.} {\bf D41} (1990) 1231-1240.

\bibitem{Ted} T.\ Frankel, {\em The Geometry of Physics}
  -- An Introduction. Cambridge University Press, Cambridge (1997).

\bibitem{Fukui} T. Fukui, {\em Propagation condition in metric-free  
electrodynamics}, Pre\-print, University of Cologne (Nov.\ 1999). 
 
\bibitem{Harnett1} G. Harnett, {\em Metrics and dual operators}, {\sl
J. Math. Phys.} {\bf 32} (1991) 84-91.

\bibitem{HLaem} M. Haugan and C. L\"ammerzahl, On the experimental
foundations of the Maxwell equations, {\sl Ann. Physik (Leipzig)} {\bf
11} (2000) this issue.

\bibitem{HOR} F.W.\ Hehl, Yu.N.\ Obukhov, and G.F.\ Rubilar, {\em
Classical electrodynamics: A Tutorial on its Foundations}. In {\sl Quo
vadis geodesia...? Festschrift for Erik W.\ Grafarend}, F.\ Krumm and
V.S.\ Schwarze (eds.) Univ.\ Stuttgart, ISSN 0933-2839 (1999) pp.\
171-184.  Los Alamos Eprint Archive, physics/9907046.
 
 
\bibitem{tHooft91} G.\ 't Hooft, {\em A chiral alternative to the
vierbein field in General Relativity}, {\sl Nucl. Phys.} {\bf B357}
(1991) 211-221.

\bibitem{Jadczyk} A.Z. Jadczyk, {\em Electromagnetic permeability of
the vacuum and light-cone structure}, {\sl Bull. Acad. Pol. Sci.,
S\'er. sci. phys. et astr.} {\bf 27} (1979) 91-94.
 
\bibitem{Laemm} C. L\"ammerzahl et al., {\em Reasons for the
electromagnetic field to obey the Maxwell equations}, Preprint,
University of Konstanz (1998).
 
\bibitem{Claus} C. L\"ammerzahl, R.A.Puntigam, and F.W.Hehl, {\em Can
  the electromagnetic field couple to post-Riemannian structures?} In
  {\sl Proceedings of the 8th Marcel Grossmann Meeting on General
  Relativity, Jerusalem 1997}, T.Piran and R.Ruffini, eds., World
  Scientific, Singapore (1999).

\bibitem{Moody} J.E. Moody and F. Wilczek, {\em New macroscopic
forces?} {\sl Phys. Rev.} {\bf D30} (1984) 130-138.

\bibitem{Ni73} W.-T. Ni, {\em A non-metric theory of gravity}.
Dept. Physics, Montana State University, Bozeman. Preprint December
1973.  [This paper is referred to by W.-T. Ni in in
{\sl Bull. Am. Phys. Soc.} {\bf 19} (1974) 655.]

\bibitem{Ni77} W.-T. Ni, {\em Equivalence principles and
electromagnetism}, {\sl Phys. Rev. Lett.} {\bf 38} (1977) 301-304.
 
\bibitem{Ni99} W.-T. Ni {\em et al.}, {\em Search for an axionlike spin
coupling using a paramagnetic salt with a dc squid}, {\sl Phys. Rev.
Lett.} {\bf 82} (1999) 2439-2442.

\bibitem{OH} Y.N.\ Obukhov and F.W.\ Hehl, {\em Space-time metric from
linear electrodynamics}, {\sl Phys. Lett.} {\bf B458} (1999) 466-470.
  
\bibitem{Tert} Yu.N. Obukhov and S.I. Tertychniy, {\it Vacuum Einstein
equations in terms of curvature forms}, {\sl Class. Quantum Grav.}
{\bf 13} (1996) 1623-1640.

\bibitem{Peres} A. Peres, {\it Electromagnetism, geometry, and the
equivalence principle}, {\sl Ann. Phys. (NY)} {\bf 19} (1962) 279-286.

\bibitem{Piron75} C. Piron, {\em \'Electrodynamique et optique, Partie
I.} Lecture Notes (edited by E. Pittet), University of Geneva (1975).

\bibitem{Piron95} C. Piron and D.J. Moore, {\em New aspects of field
theory}, {\sl Turk. J. Phys.} {\bf 19} (1995) 202-216.

\bibitem{Post95} E.J. Post, {\em Quantum Reprogramming -- Ensembles
and Single Systems: A Two-Tier Approach to Quantum Mechanics,}
Dordrecht, Kluwer (1995).

\bibitem{PLH} R.A.\ Puntigam, C.\ L\"ammerzahl, and F.W.\ Hehl, {\em
Maxwell's theory on a post--Riemannian spacetime and the equivalence
principle}, {\sl Class. Quantum Grav.} {\bf 14} (1997) 1347-1356.
 
\bibitem{Schoenberg} M.\ Sch\"onberg, {\em Electromagnetism and
gravitation}, {\sl Rivista Brasileira de Fisica} {\bf 1} (1971)
91-122.
 
\bibitem{SU} R.U.\ Sexl and H.K.\ Urbantke, {\em Gravitation und
Kosmologie: Eine Einf\"uhrung in die Allgemeine
Relativit\"atstheorie}, 4th rev. ed., Spektrum Akad. Verlag,
Heidelberg (1995) p.\ 50.
 
\bibitem{Toupin} R.A. Toupin, {\em Elasticity and
electro-magnetics}. In {\sl Non-Linear Continuum Theories, C.I.M.E.\
Conference, Bressanone, Italy (1965)}. C. Truesdell and G. Grioli
coordinators. Pp.\ 206-342.

\bibitem{Urban1} H. Urbantke, {\em A quasi-metric associated with
$SU(2)$ Yang-Mills field}, {\sl Acta Phys. Austriaca Suppl.} {\bf 29}
(1978) 875-816.
 
\bibitem{Urban2} H. Urbantke, {\em On integrability properties of
$SU(2)$ Yang-Mills fields. I. Infinitesimal part}, {\sl J. Math.
Phys.} {\bf 25} (1984) 2321-2324.

\bibitem{Wang} C. Wang, {\em Mathematical Principles of Mechanics and 
Electromagnetism, Part B: Electromagnetism and Gravitation}, Plenum 
Press, New York (1979). 
 
\bibitem{Weinberg} S. Weinberg, {\em A new light boson?} {\sl Phys.
Rev. Lett.} {\bf 40} (1978) 223-226.

\bibitem{Wilczek1} F. Wilczek, {\em Problem of strong P and T
invariance in the presence of instantons}, {\sl Phys. Rev. Lett.} {\bf
40} (1978) 279-282.
 
\end{thebibliography}
\end{document}